\documentclass[aps,pre,twocolumn,floatfix,showpacs]{revtex4}
\usepackage{graphicx}
\usepackage{bm}
\bibstyle{apsrev.bst}

\newcommand{\ave}[1]{\langle #1\rangle}

\begin{document}
\title{Comment on: ``Solving the riddle of the bright mismatches: Labeling and effective
binding in oligonucleotide arrays"}
\author{E. Carlon}
\affiliation{Interdisciplinary Research Institute c/o IEMN, Cit\'e
Scientifique BP 60069, F-59652 Villeneuve d'Ascq, France}
\author{T. Heim}
\affiliation{Interdisciplinary Research Institute c/o IEMN, Cit\'e
Scientifique BP 60069, F-59652 Villeneuve d'Ascq, France}
\author{J.~Klein Wolterink}
\affiliation{Institute for Theoretical Physics, University of Utrecht,
Leuvenlaan 4, 3584 CE Utrecht}
\author{G.~T.~Barkema}
\affiliation{Institute for Theoretical Physics, University of Utrecht,
Leuvenlaan 4, 3584 CE Utrecht}

\date{\today}

\begin{abstract}
In a recent paper [Phys. Rev. E {\bf 68}, 011906 (2003)], Naef
and Magnasco suggested that the ``bright" mismatches observed in
Affymetrix microarray experiments are caused by the fluorescent
molecules used to label RNA target sequences, which would impede
target-probe hybridization. Their conclusion is based on the observation
of ``unexpected" asymmetries in the affinities obtained by fitting
microarray data from publicly available experiments.  We point out
here that the observed asymmetry is due to the inequivalence of RNA and DNA,
and that the reported affinities are consistent with stacking free
energies obtained from melting experiments of {\it unlabeled} nucleic
acids in solution.  The conclusion of Naef and Magnasco is therefore
based on an unjustified assumption.
\end{abstract}


\maketitle

\newcommand{\ul}{\underline}
\newcommand{\bc}{\begin{center}}
\newcommand{\ec}{\end{center}}
\newcommand{\be}{\begin{equation}}
\newcommand{\ee}{\end{equation}}
\newcommand{\ba}{\begin{array}}
\newcommand{\ea}{\end{array}}
\newcommand{\beqn}{\begin{eqnarray}}
\newcommand{\eeqn}{\end{eqnarray}}

In a recent paper \cite{naef03} Naef and Magnasco investigated the problem
of bright mismatches (MMs) in Affymetrix DNA microarrays.  In these
arrays a perfect matching (PM) 25 nucleotides probe is accompanied by a
MM one in which a single nucleotide in the central position is modified.
The MM probes in Affymetrix chips are introduced with the purpose of
estimating the contributions from non-specific hybridization.  One of
the problems with this approach is that of the so-called ``bright"
mismatches, for which the fluorescence intensity measured from a MM
probe is higher than that from the corresponding PM probe.  This is
seemingly at odds with basic thermodynamics, as a perfectly matching
duplex is more stable than one containing a mismatch.  The analysis
of experimental data shows that the occurrence of bright mismatches is
rather frequent. It was found to occur in $30\%$ of the probes in the
Affymetrix Human HGU95a microarray \cite{naef02}.

The authors of Ref. \cite{naef03} analyze a series of microarray
experiments performed by Affymetrix.  In this set of experiments the
targets are single stranded RNA molecules where some of the pyrimidines
(the U and C bases) carry a biotin, while the probes are single
stranded DNA molecules.  After the hybridization step is terminated
the solution containing the non-hybridized target is washed off from
the array. Fluorescent labels, attached to streptadivin molecules, are
then added. The biotin is a strong binding site for the streptadivin
during this staining step. Although Naef and Magnasco use ``biotin" and
``fluorescent label" as synonyms in Ref. \cite{naef03}, it is important
to emphasize that there are no fluorescent labels during the hybridization
step in Affymetrix experiments.

%%%%%%%%%%%%%%%%%%%%%%%%%%%%%%%%%%% TAB.I %%%%%%%%%%%%%%%%%%%%%%%%%%%%%%%%%
\begin{table}[t]
\caption{The stacking free energy parameters $\Delta G_{\rm 37}^\circ$
for RNA/DNA hybrids measured in solution at a salt concentration $1$
M NaCl and $T=37^\circ$ \cite{sugi95_sh}. The upper strand is RNA (with
orientation 5'-3') and lower strand DNA (orientation 3'-5').}
\begin{ruledtabular}
\begin{tabular}{cc|cc}
Sequence & $-\Delta G_{\rm 37}$(kcal/mol)  & Sequence & $-\Delta G_{\rm 37}$(kcal/mol)\\
\hline
&&&\\
${\rm rAA} \atop {\rm dTT}$ & 1.0 &
${\rm rAC} \atop {\rm dTG}$ & 2.1 \\
&&&\\
${\rm rAG} \atop {\rm dTC}$ & 1.8 &
${\rm rAU} \atop {\rm dTA}$ & 0.9 \\
&&&\\
${\rm rCA} \atop {\rm dGT}$ & 0.9 &
${\rm rCC} \atop {\rm dGG}$ & 2.1 \\
&&&\\
${\rm rCG} \atop {\rm dGC}$ & 1.7 &
${\rm rCU} \atop {\rm dGA}$ & 0.9 \\
&&&\\
${\rm rGA} \atop {\rm dCT}$ & 1.3 &
${\rm rGC} \atop {\rm dCG}$ & 2.7 \\
&&&\\
${\rm rGG} \atop {\rm dCC}$ & 2.9 &
${\rm rGU} \atop {\rm dCA}$ & 1.1 \\
&&&\\
${\rm rUA} \atop {\rm dAT}$ & 0.6 &
${\rm rUC} \atop {\rm dAG}$ & 1.5 \\
&&&\\
${\rm rUG} \atop {\rm dAC}$ & 1.6 &
${\rm rUU} \atop {\rm dAA}$ & 0.2 \\
&&&
% T=37
% -0.20 -1.60 -1.50 -0.60 -1.10 -2.90 -2.70 -1.30 
% -0.90 -1.70 -2.10 -0.90 -0.90 -1.80 -2.10 -1.00
% T=45 
%  0.08 -1.37 -1.31 -0.42 -0.93 -2.65 -2.56 -1.21
% -0.73 -1.32 -1.92 -0.70 -0.70 -1.62 -1.99 -0.83
% 
\end{tabular}
\end{ruledtabular}
\label{table_DG}
\end{table}
%%%%%%%%%%%%%%%%%%%%%%%%%%%%%%%%%%%%% TAB.I %%%%%%%%%%%%%%%%%%%%%%%%%%%%%%%%%

In Ref. \cite{naef03} the brightness of PM probes is fitted using
affinities $A_{l,i}$ in which $l=A, C, T$ or $G$ denotes the base type
and $i=1,2 \ldots 25$ the position along the probe.  The resulting
fit-values are plotted in their Fig. 3.  Notably, there are differences
between affinities for A,T and C,G bases, which were interpreted in
Ref. \cite{naef03} as as due to the presence of biotin in some bases
of the RNA strand.  We quote from Ref. \cite{naef03}: `` \ldots {\it An
unexpected aspect of the above fits is the asymmetry of A versus T (and
G versus C) affinities, which goes against the zeroth order energetic
consideration that A-T and T-A bonds (or G-C and C-G) would contribute
equally to the binding \ldots The obvious culprits are the fluorescent
labels \ldots}".

We strongly disagree with this interpretation of the data.  There is no
``symmetry'' between A-T and T-A, since one nucleotide is part of an
RNA strand and the other of DNA. To start with, RNA does not have a T
(thymine) but a U (uracil), and the authors compare an A-T binding with
a U-A binding; and although the usual naming of a C-G and a G-C binding
suggests symmetry, also this symmetry is broken by the different backbones
of RNA and DNA.

% %%% ADDED TEXT
% The thermodynamics of RNA/DNA duplexes in solution has been investigated
% in a series of experiments in which the melting temperatures
% of short duplexes, of about 20 nucleotides, are measured (see eg
% \cite{sugi95_sh}). These measurements provide estimates of $\Delta H$
% and $\Delta S$ the enthalpy and entropy differences between a duplex and
% the two separate strands.  As for other types of duplexes, eg DNA/DNA
% and RNA/RNA, it turns out that $\Delta H$ and $\Delta S$ can be rather
% well approximated by sums of sequence dependent local terms taking into
% account the contribution of hydrogen bondings and stacking interactions
% between neighboring bases. As a consequence the free energy difference,
% or hybridization free energy, $\Delta G = \Delta H - T \Delta S$ can
% also be approximated by a sum of local terms.

The thermodynamics of RNA/DNA duplexes in solution has been investigated
in a series of experiments in which the melting temperatures of
short duplexes, of about 20 nucleotides, are measured (see e.g.
\cite{sugi95_sh}). These measurements provide estimates of the differences
in enthalpy $\Delta H$ and entropy $\Delta S$ between a duplex and the
two separate strands.  As for other types of duplexes, e.g. DNA/DNA
and RNA/RNA, it turns out that $\Delta H$ and $\Delta S$ can be well
approximated by sums of sequence dependent local terms taking into
account the contribution of hydrogen bonds and stacking interactions
between neighboring bases. As a consequence the free energy difference,
or hybridization free energy, $\Delta G = \Delta H - T \Delta S$ is also
well approximated by a sum of local terms.
%%%%%%%%%% 
The latter are given in Table \ref{table_DG} \cite{sugi95_sh}.  Note the
asymmetries between the free energy parameters when interchanging
nucleotides between DNA and RNA strands.  

In order to compare the hybridization free energies in solution with the
affinities reported in Ref. \cite{naef03}, which are single nucleotide
dependent, we fix a nucleotide on the probe strand and average the
values in Table \ref{table_DG} over all possible neighbors in the 3'
and 5' direction.  For instance for a T in the DNA strand we define:
\beqn
\Delta G_{\rm T}
\equiv \frac 1 4
\sum_{\gamma \in \{A,T,G,C\} }
\left[
\Delta G \left( {{\rm rA\gamma'} \atop {\rm dT\gamma}} \right) +
\Delta G \left( {{\rm r\gamma'A} \atop {\rm d\gamma T}} \right)
\right]
\label{average}
\eeqn
where $\gamma'$ is the nucleotide in the RNA strand complementary to
$\gamma$. These free energies and the corresponding binding affinities
are given in Table \ref{tableII}.

%%%%%%%%%%%%%%%%%%%%%%%%%%%%%%%%%%%%%%%%%%%%%%%%%%%%%%%%%%%%%%%%%%%%%%%%%%%
\begin{table}[t]
\caption{
Column 2: Single nucleotide free energies obtained from Eq. (\ref{average})
expressed in kcal/mole. Column 3: The same free energies to which the average
value is subtracted. Column 4: The $\log 10$ affinities derived from the
data of column 3 by $A_\gamma = (\langle \Delta G \rangle - \Delta
G_\gamma)/(RT \ln 10)$,  with $R T = 0.63$ kcal/mole.
Column 5: Effective affinities for the middle bases as given in the Fig. 3
of Ref. \cite{naef03}.
}
\label{tableII}
\begin{ruledtabular}
\begin{tabular}{ccccc}
$\gamma$ & $-\Delta G_\gamma$ & $\ave{\Delta G} - \Delta G_\gamma$ & $A_\gamma$ &
Ref. \cite{naef03} \\
\hline
 C &  4.00 &  1.09 &  0.75 & 0.20 \\
 G &  3.50 &  0.59 &  0.40 & 0.02 \\
 T &  2.40 & -0.51 & -0.35 &-0.01 \\
 A &  1.75 & -1.16 & -0.80 &-0.20
\end{tabular}
\end{ruledtabular}
\end{table}
%%%%%%%%%%%%%%%%%%%%%%%%%%%%%%%%%%%%%%%%%%%%%%%%%%%%%%%%%%%%%%%%%%%%%%%%%%%
 
Note that Eq. (\ref{average}) gives lower affinities for A compared to
T and for G compared to C, in qualitative agreement with the data of
Ref. \cite{naef03} in Table \ref{tableII}. While in Ref. \cite{naef03}
these differences were argued to prove that biotin affects the binding,
our analysis clearly shows that these differences (or asymmetries as
referred to in \cite{naef03}) are intrinsic properties of unbiotinylated
RNA/DNA duplexes in solution.

It is not surprising that the effective affinities measured in
Ref.~\cite{naef03} are smaller than the binding free energies obtained
from Eq. (\ref{average}). The affinities of Ref. \cite{naef03} are
obtained by fitting the measured fluorescent signals of the microarray to
a Langmuir model (Eq. (1) in \cite{naef03}). The fluorescence measured
in the microarray experiment is not solely determined by the binding
free energy between an isolated probe and a specific target, but it is
also influenced by many other effects like polydispersity in probe and
target lengths, secondary structure formation in probes and targets,
and hybridization between targets in solution.  We thus do not expect
that the affinities of Ref. \cite{naef03} should agree quantitatively
with the binding free energies in solution.  As the neglected processes
compete with the hybridization of a probe with a complementary target,
it is to be expected that the difference in effective affinities of
Ref. \cite{naef03} are lower than their solution counterparts.

Certainly, Ref. \cite{naef03} does not show that fluorescent labels
(or, to be precise, the biotin linker) interfere with binding, or are
the cause of bright mismatches.

We acknowledge financial support from the Van Gogh Programme d'Actions
Int\'egr\'ees (PAI) 08505PB of the French Ministry of Foreign Affairs
and NWO grant 62403735.

% \bibliography{biblio.bib}
% \bibliography{/home/enrico/TEX/biblio.bib}

\end{document}